\documentclass[journal]{IEEEtran}

\ifCLASSINFOpdf
\else
   \usepackage[dvips]{graphicx}
\fi
\usepackage{url}
\usepackage{listings}
\usepackage{caption}
\usepackage{tabularx}
\usepackage{tikz}
\usetikzlibrary{arrows.meta}

\usepackage{amsmath,bm, amssymb, setspace}
\usepackage{multirow,bigdelim}

\usepackage{makecell}
\usepackage{xcolor,colortbl}
\usepackage{color}
\usepackage{colortbl}
\usepackage{hyperref}
\usepackage[shortlabels]{enumitem}
\newcommand{\zhenk}[1]{{\color{black}{#1}}}

\usepackage[pscoord]{eso-pic}

\newcommand{\placetextbox}[3]{
  \setbox0=\hbox{#3}
  \AddToShipoutPictureFG*{
    \put(\LenToUnit{#1\paperwidth},\LenToUnit{#2\paperheight}){\vtop{{\null}\makebox[0pt][c]{#3}}}%
  }%
}%

\ifCLASSOPTIONcompsoc
  \usepackage[nocompress]{cite}
\else
  \usepackage{cite}
\fi
\ifCLASSINFOpdf
\else
\fi
\usepackage[]{footmisc}

\definecolor{creamson}{RGB}{153, 0, 0}

\usepackage{graphicx}

\usepackage{subfigure}
\input{def.set}

\begin{document}

\title{
Psychoacoustic Calibration of Loss Functions for Efficient End-to-End Neural Audio Coding
}

\author{Kai Zhen, \IEEEmembership{Student Member, IEEE}, Mi Suk Lee, Jongmo Sung, Seungkwon Beack,\\Minje Kim, \IEEEmembership{Senior Member, IEEE}
\thanks{This work was supported by the Institute for Information and Communications Technology Promotion (IITP) funded by the Korea government (MSIT) under Grant 2017-0-00072 (Development of Audio/Video Coding and Light Field Media Fundamental Technologies for Ultra Realistic Tera-Media).

Kai Zhen is with the Department of Computer Science and Cognitive Science Program at Indiana University, Bloomington, IN 47408 USA. Mi Suk Lee, Jongmo Sung, and Seungkwon Beack are with Electronics and Telecommunications Research Institute, Daejeon, Korea 34129. Minje Kim is with the Dept. of Intelligent Systems Engineering at Indiana University (e-mails: zhenk@iu.edu, lms@etri.re.kr, jmseong@etri.re.kr, skbeack@etri.re.kr, minje@indiana.edu).}}

\maketitle

\begin{abstract}
Conventional audio coding technologies commonly leverage human perception of sound, or psychoacoustics, to reduce the bitrate while preserving the perceptual quality of the decoded audio signals. For neural audio codecs, however, the objective nature of the loss function usually leads to suboptimal sound quality as well as high run-time complexity due to the large model size. 
In this work, we present a psychoacoustic calibration scheme to re-define the loss functions of neural audio coding systems so that it can decode signals more perceptually similar to the reference, yet with a much lower model complexity.
The proposed loss function incorporates the global masking threshold, allowing the reconstruction error that corresponds to inaudible artifacts.
Experimental results show that the proposed model outperforms the baseline neural codec twice as large and consuming 23.4\% more bits per second. 
With the proposed method, a lightweight neural codec, with only 0.9 million parameters, performs near-transparent audio coding comparable with the commercial MPEG-1 Audio Layer III codec at 112 kbps.
\end{abstract}

\begin{IEEEkeywords}
Audio coding, deep neural networks, psychoacoustics, network compression
\end{IEEEkeywords}

\IEEEpeerreviewmaketitle


\placetextbox{0.5}{1}{\scriptsize \textit{The article has been accepted for publication by IEEE. Full citation:}}
\placetextbox{0.5}{.99}{\scriptsize Kai Zhen, Mi Suk. Lee, Jongmo Sung, Seungkwon Beack and Minje Kim, ``Psychoacoustic Calibration of Loss Functions for Efficient End-to-End Neural Audio Coding,"}
\placetextbox{0.5}{.98}{\scriptsize in IEEE Signal Processing Letters, vol. 27, pp. 2159-2163, 2020.}
\placetextbox{0.5}{.97}{\scriptsize DOI: 10.1109/LSP.2020.3039765.}
\placetextbox{0.5}{.03}{\scriptsize 
\copyright 2020 IEEE. Personal use of this material is permitted. Permission from IEEE must be obtained for all other uses, in any current or future media, including reprinting/republishing this material }
\placetextbox{0.5}{.02}{\scriptsize for advertising or promotional purposes, creating new collective works, for resale or redistribution to servers or lists, or reuse of any copyrighted component of this work in other works.}

\section{Introduction}

\IEEEPARstart{A}{udio} 
coding, a fundamental set of technologies in data storage and communication, compresses the original signal into a bitstream with a minimal bitrate (encoding) without sacrificing the perceptual quality of the recovered waveform (decoding) \cite{BosiM1997iso, mp3}. 
In this paper we focus on the lossy codecs, which typically allow information loss during the process of encoding and decoding only in inaudible audio components. To this end, psychoacoustics is employed to quantify the audibility in both time and frequency domains.
For example, MPEG-1 Audio Layer III (also known as MP3), as a successful commercial audio codec, achieves a near-transparent quality at 128 kbps by using a psychoacoustic model (PAM) \cite{mp3}. Its bit allocation scheme determines the number of bits allocated to each subband by dynamically computing the masking threshold via a PAM and then allowing quantization error once it is under the threshold \cite{BrandenburgK1994iso}. 

Recent efforts on deep neural network-based speech coding systems have made substantial progress on the coding gain \cite{engel2017neural, OordA2016wavenet, KlejsaJ2019samplernn}.
They formulate coding as a complex learning process that converts an input to a compact hidden representation. This poses concerns for edge applications with the computational resource at a premium: a basic U-Net audio codec contains approximately 10 million parameters \cite{StollerD2018waveunet}; in \cite{GarbaceaC2019vqvae}, vector quantized variational autoencoders (VQ-VAE) \cite{OordA2017vqvae} employs WaveNet \cite{OordA2016wavenet} as a decoder, yielding a competitive speech quality at 1.6 kbps, but with 20 million parameters.
In addition, recent neural speech synthesizers employ traditional DSP techniques, e.g., linear predictive coding (LPC), to reduce its complexity \cite{ValinJ2019lpcnet}. Although it can serve as a decoder of a speech codec, LPC does not generalize well to non-speech signals.

Perceptually meaningful objective functions have shown an improved trade-off between performance and efficiency. \zhenk{Some recent speech enhancement models successfully employed perceptually inspired objective metrics, e.g. perceptual attractors \cite{ChenZ2017deep}, energy-based weighting \cite{LiuQ2017perceptually}, perceptual weighting filters from speech coding \cite{zhao2019perceptual}, and global masking thresholds \cite{kumar2016speech} \cite{ZhenK2018psychoacoustically}, while they have not targeted audio coding and model compression.} Other neural speech enhancement systems implement short-time objective intelligibility (STOI) \cite{TaalC2010icassp} and perceptual evaluation of speech quality (PESQ) \cite{RixA2001pesq} as the loss \cite{FuS2018e2e, MartinJ2018deep}. These metrics may benefit speech codecs, but do not faithfully correlate with subjective audio quality. 
Meanwhile, PAM serves as a subjectively salient quantifier for the sound quality and is pervasively used in the standard audio codecs. However, integrating the prior knowledge from PAM into optimizing neural audio codecs has not been explored.

In this paper, we present a psychoacoustic calibration scheme to improve the neural network optimization process, as an attempt towards efficient and high-fidelity neural audio coding (NAC). With the global masking threshold calculated from a well-known PAM \cite{PainterT2000ieeeproc}, the scheme firstly conducts priority weighting making the optimization process focus more on audible coding artifacts in frequency subbands with the relatively weaker masking effect, while going easy otherwise. The scheme additionally modulates the coding artifact to ensure that it is below the global masking threshold, which is analogous to the bit allocation algorithm in MP3 \cite{mp3}. 
This is, to our best knowledge, the first method to directly incorporate psychoacoustics to neural audio coding.

\begin{figure}[t]
    \centering
     \includegraphics[width=\columnwidth]{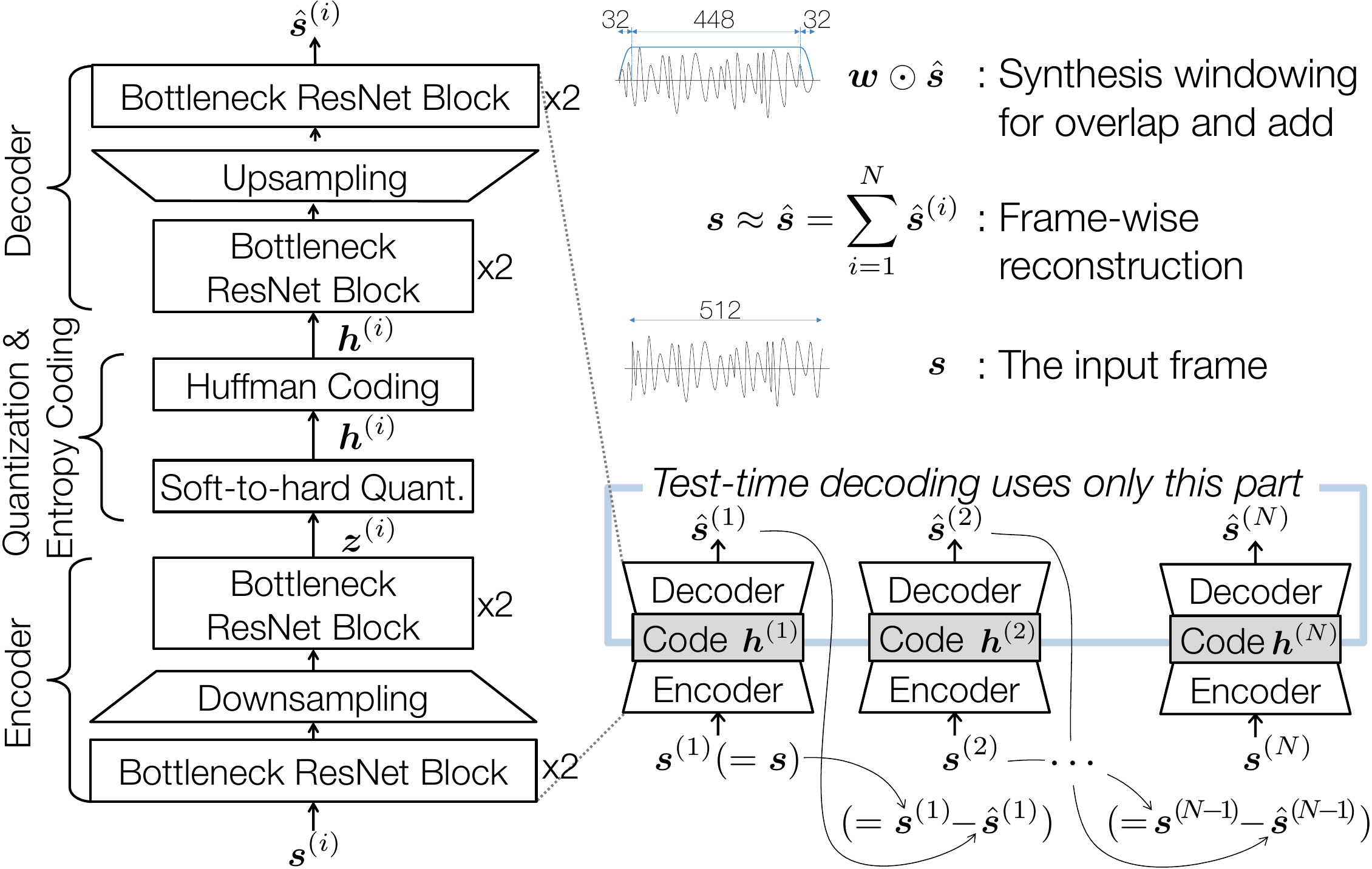}
    \vspace{-0.1in}
    \caption{Schematic diagrams for NAC. The residual coding pipeline for CMRL consists of multiple NAC autoencoding modules. Training and test-time encoding uses all blocks while the test-time decoding uses only the decoder portion.}\vspace{-0.1in}
    \label{fig:nac}
\end{figure}

\section{End-to-end neural audio coding}
\label{sec:pre}

\subsection{Lightweight NAC Module}
\label{sec:nac}

Given that neural codecs can suffer from a large inference cost due to their high model complexity, one of our goals is to demonstrate the advantage of the proposed pychoacoustic loss function on model compression. To that end, we choose a compact neural audio coding (NAC) module as the building block.
The NAC module is a simplified version of a convolutional neural network (CNN)-based autoencoder \cite{KankanahalliS2018icassp} with only 450K parameters. As shown in Fig. \ref{fig:nac}, it consists of a stack of bottleneck blocks as in \cite{TanK2019taslp}, each of which performs a ResNet-style residual coding \cite{HeK2016cvpr}. The code vector produced by its encoder part is discretized into a bitstring via the soft-to-hard quantization process originally proposed in \cite{AgustssonE2017softmax} for image compression. We detail the description as follows.

\subsubsection{Encoder}
The CNN encoder maps an input frame of $T$ time-domain samples, $\bs\in\Real^T$ to the code vector, i.e., $\bz\leftarrow \mathcal{F}_\text{enc}(\bs)$. Striding during the 1D convolution operation can downsample the feature map. For example,  $\bz\in\Real^{T/2}$ when the stride is set to be 2 and applied once during encoding. The detailed architecture is summarized in TABLE \ref{tab:topo}.

\subsubsection{Soft-to-hard quantization}
Quantization replaces each real-valued element of the code vector $\bz$ with a kernel value chosen from a set of $K$ representatives.
We use soft-to-hard quantizer \cite{AgustssonE2017softmax}, a clustering algorithm compatible with neural networks, where the representatives are also trainable. During training, in each feedforward routine, the $c$-th code value $z_c$ is assigned to the nearest kernel out of $K$, $\bbeta\in\Real^K$, which have been trained so far. 
The discrepancy between $z_c$ and the chosen kernel $h_c \in \{\beta_1, \beta_2, \cdots, \beta_K\}$ (namely the quantization error) is accumulated in the final loss, and then reduced during training via backpropagation (i.e., by updating the means and assignments). 
Specifically, the cluster assignment is conducted  by calculating the distance, $\bd\in\Real^K$, between the code value and all kernels, and then applying the softmax function to the negatively scaled distance to produce a probabilistic membership assignment: $\ba\leftarrow\text{softmax}(-\alpha\bd)$. Although we eventually need a hard assignment vector $\bsa$, i.e., a one-hot vector that indicates the closest kernel, during training the quantized code $\bh$ is acquired by a soft assignment, $\ba^\top\bbeta$, for differentiability. Hence, at the test time, $\bsa$ replaces $\ba$ by turning on only the maximum element. Note that a larger scaling factor $\alpha$ makes $\ba$ harder, making it more similar to $\bsa$. Huffman coding follows to  generate the final bitstream.

\subsubsection{Decoder}
The decoder recovers the original signal from the quantized code vector:  $\hat\bs=\mathcal{F}_{\text{dec}}(\bh)$, by using an architecture mirroring that of the encoder (TABLE \ref{tab:topo}). For upsampling, we use a sub-pixel convolutional layer proposed in \cite{ShiW2016superresolution} to recover the original frame length $T$.

\begin{table}[t]
\centering
\caption{
The 1D-CNN NAC module architecture (Fig.~\ref{fig:nac}). The shape of feature maps is (frame length, channel); the kernel shape is (kernel size, in channel, out channel).
}
\label{tab:topo}
\setlength\tabcolsep{2.6pt}
\footnotesize
\begin{tabular}{ |c|c|c|c|c| }
 \hline
 System & Layer &Input shape & Kernel shape & Output shape\\
 \hline
\multirow{10}{*}{Encoder} & Change channel & (512, 1) & (9, 1, 100) &(512, 100) \\
\cline{2-5}
& 1st bottleneck & (512, 100) & \begin{tabular}{cc}\rule[6pt]{0pt}{0pt}(9, 100, 20)  &\rdelim]{2.8}{5mm}[$\times$2]\\ (9, 20, 20)  &  \\(9, 20, 100) &\rule[-2pt]{0pt}{0pt}\end{tabular} &(512, 100)  \\\cline{2-5}
& Downsampling & (512, 100) & (9, 100, 100) &(256, 100) \\
\cline{2-5}
& 2nd bottleneck & (256, 100) & \begin{tabular}{cc}\rule[6pt]{0pt}{0pt}(9, 100, 20)  &\rdelim]{2.8}{5mm}[$\times$2]\\ (9, 20, 20)  &  \\(9, 20, 100) &\rule[-2pt]{0pt}{0pt}\end{tabular} &(256, 100)  \\
\cline{2-5}
& Change channel & (256, 100) & (9, 100, 1)& (256, 1) \\
\hline
\multicolumn{5}{|c|}{Soft-to-hard quantization \& Huffman coding}\\
\hline
\multirow{10}{*}{Decoder}& Change channel & (256, 1) & (9, 1, 100) &(256, 100) \\
\cline{2-5}
& 1st bottleneck & (256, 100) & \begin{tabular}{cc}\rule[6pt]{0pt}{0pt}(9, 100, 20)  &\rdelim]{2.8}{5mm}[$\times$2]\\ (9, 20, 20)  &  \\(9, 20, 100) &\rule[-2pt]{0pt}{0pt}\end{tabular} &(256, 100)  \\\cline{2-5}
& Upsampling & (256, 100) & (9, 100, 100) &(512, 50) \\
\cline{2-5}
& 2nd bottleneck & (512, 50) & \begin{tabular}{cc}\rule[6pt]{0pt}{0pt}(9, 50, 20)  &\rdelim]{2.8}{5mm}[$\times$2]\\ (9, 20, 20)  &  \\(9, 20, 50) &\rule[-2pt]{0pt}{0pt}\end{tabular} &(512, 50)  \\\cline{2-5}
& Change channel & (512, 50) & (9, 50, 1) & (512, 1) \\
\hline
\end{tabular}
\vspace{-.1in}
\end{table}

\zhenk{
\subsubsection{Bitrate Analysis and Control}
The lower bound of the bitrate is defined as $|\bh|\mathcal{H}(\bh) $, where $|\bh|$ is the number of down-sampled and quantized features per second. The entropy $\mathcal{H}(\bh)$ forms the lower bound of the average amount of bits per feature. 
While $|\bh|$ is a constant given a fixed sampling rate and network topology, $\mathcal{H}(\bh)$ is adaptable during training. As detailed in \cite{AgustssonE2017softmax}, basic information theory calculates $\mathcal{H}(\bh)$ as $-\sum_k p(\beta_k)\log_{2}p(\beta_k)$, where $p(\beta_k)$ denotes the occurrence probability of the $k$-th cluster defined in the soft-to-hard quantization. Therefore, during model training, $\mathcal{H}(\bh)$ is added to the loss function as a regularizer navigating the model towards the target bitrate. Initiated as $0.0$, the blending weight increases by $0.015$ if the actual bitrate overshoots the target and decreases by that amount otherwise. Because this regularizer is well defined in the literature \cite{AgustssonE2017softmax}\cite{KankanahalliS2018icassp}\cite{ZhenK2019interspeech}, we omit it in following sections for simplicity purposes.
}

\subsection{Cross-Module Residual Learning}
\label{sec:cmrl}
To scale up for high bitrates, cross-module residual learning (CMRL) \cite{ZhenK2019interspeech} implants the multistage quantization scheme \cite{GershoA1983vq} by cascading residual coding blocks (Fig. \ref{fig:nac}). CMRL decentralizes the neural autoencoding effort to a chain of serialized low complexity coding modules, with the input of $i$-th module being $\bs^{(i)} = \bs - \sum_{j=1}^{i-1}\hat{\bs}^{(j)}$. That said, each module only encodes what is not reconstructed from preceding modules, making the system scalable. Concretely, for an input signal $\bs$, the encoding process runs all $N$ autoencoder modules in a sequential order, which yields the bitstring as a concatenation of the quantized code vectors:  $\bh=\big[{\bh^{(1)}}^\top,{\bh^{(2)}}^\top,\cdots,{\bh^{(N)}}^\top\big]^\top$. 
During decoding, all decoders,  $\calF_\text{dec}(\bh^{(i)}) ~~ \forall i$,
run to produce the reconstructions that sum up to approximate the initial input signal as $\sum_{i=1}^{N}\hat{\bs}^{(i)}$. 

\section{The Proposed Psychoacoustic Calibration}
\label{sec:calibration}

The baseline model uses the sum of squared error (SSE) defined in the time domain: $\mathcal{L}_1(\bs||\hat{\bs}) = \sum_{i=1}^N \sum_{t=1}^T\big(\hat{s}_t^{(i)}-s_t^{(i)}\big)^2$. In addition, another loss is defined in the mel-scaled frequency domain to weigh more on the low frequency area, as the human auditory system does, $\mathcal{L}_2(\by||\hat{\by}) =
\sum_{i=1}^N \sum_{l=1}^L \big(y_l^{(i)}-\hat{y}_l^{(i)}\big)^2$, where $\by$ stands for a mel spectrum with $L$ frequency subbands as proposed in \cite{KankanahalliS2018icassp}. 

\subsection{Psychoacoustic Model-1}
\label{sec:pam}
\begin{figure}
   \centering
   {\includegraphics[width=.95\columnwidth]{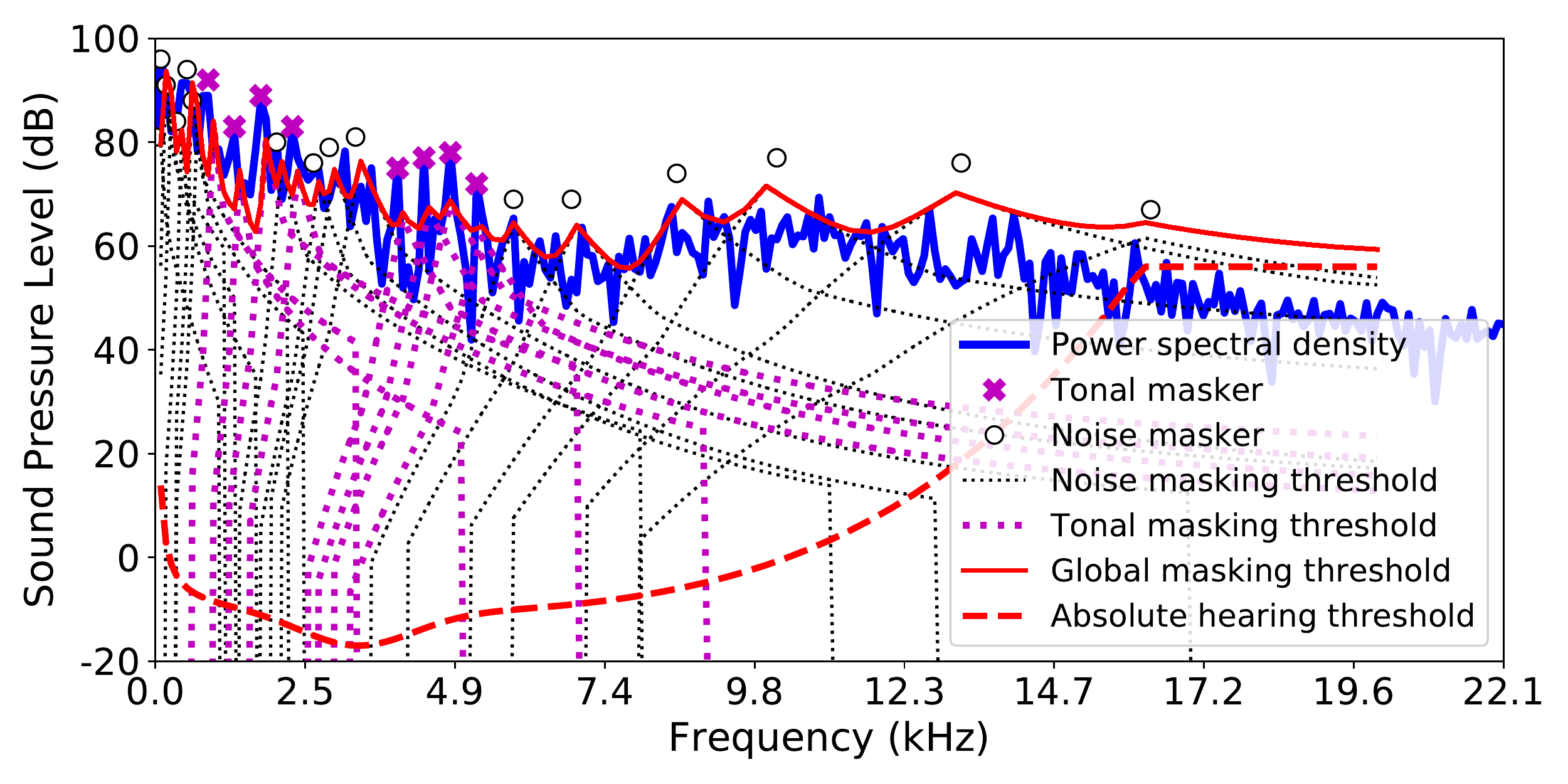}} \vspace{-0.02in}
    \caption{Visualization of the masker detection, individual and global masking threshold calculation for an audio input.}
    \vspace{-0.05in}
    \label{fig:pam}
\end{figure}

Without loss of generality, we choose a basic PAM that computes simultatenous masking effects for the input signal as a function of frequency, while the temporal masking effect is not integrated. According to PAM-1 defined in \cite{mp3}, for an input frame, it (a) calculates the logarithmic power spectral density (PSD) $\bp$; (b) detects tonal and noise maskers, followed by decimation; \zhenk{(c) calculates masking threshold for individual tonal and noise maskers $\bU\in\Real^{F\times R}, \bV\in\Real^{F\times B}$, where $R$ and $B$ are the number of maskers. The global masking threshold at frequency bin $f$ is accumulated from each individual masker in (c) along with the absolute hearing threshold $\bQ$ \cite{PainterT2000ieeeproc}, as 
$m_f = 10\log_{10}\left( 10^{0.1 Q_f} + \sum_r 10^{0.1 U_{f,r}} + \sum_b 10^{0.1 V_{f,b}} \right)$}.
Fig.~\ref{fig:pam} shows an example of $\bp$ of a signal and its global masking threshold based on the simultaneous masking effect.

Global masking threshold as discussed is used in various conventional audio codecs to allocate minimal amount of bits without losing the perceptual audio quality.
Typically, the bit allocation algorithm optimizes $n_f/m_f$ (NMR), where $n_f$ denotes the power of the noise (i.e., coding artifacts) in the subband $f$ and $m_f$ is the power of the global masking threshold. In an iterative process, each time the bit is assigned to the subband with the highest NMR until no more bit can be allocated \cite{salomon2004data, yen2005low, zamani2019spatial}.
\zhenk{The global masking curve acquired via PAM-1 comprises both input-invariant prior knowledge as in the absolute hearing threshold and input-dependent masking effects. We propose two mechanisms to integrate PAM-1 into NAC optimization: priority weighting and noise modulation. }

\subsection{Priority Weighting}

During training we estimate the logarithmic PSD $\bp$ out of an input frame $\bs$, as well the global masking threshold $\boldm$ to define a perceptual weight vector, $\bw = \log_{10}(\frac{10^{0.1\bp}}{10^{0.1\boldm}}+1)$: the log ratio between the signal power and the masked threshold, rescaled from decibel. Accordingly, we define a weighting scheme that pays more attention to the unmasked frequencies: 
\begin{equation}
\mathcal{L}_3(\bs||\hat{\bs}) = \sum_i\sum_f w_f\Big(x_f^{(i)}-\hat{x}_f^{(i)}\Big)^2,  
\label{eq:SMR}
\end{equation}
where $x_f^{(i)}$ and $\hat{x}_f^{(i)}$ are the $f$-th magnitude of the Fourier spectra of the input and the recovered signals for the $i$-th CMRL module. The intuition is that, if the signal’s power is greater than its masking threshold at the $f$-th frequency bin, i.e. $p_f > m_f$, the model tries hard to recover this audible tone precisely: a large $w_f$ enforces it. Otherwise, for a masked tone, the model is allowed to generate some reconstruction error. The weights are bounded between 0 and $\infty$, 
whose smaller extreme happens if, for example, the masking threshold is too large comparing to the sufficiently soft signal.

\subsection{Noise Modulation}

\zhenk{The priority weighting mechanism can accidentally result in audible reconstruction noise, exceeding the mask value $m_f$, when $w_f$ is small. Our second psychoacoustic loss term is to modulate the reconstruction noise by directly exploiting NMR, 
$n_f/m_f$, where $\bn$ is the power spectrum of the reconstruction error $\bs-\sum_{i=1}^{N}\hat{\bs}^{(i)}$ from all $N$ autoencoding modules. 
We tweak the greedy bit allocation process in the MP3 encoder that minimizes NMR iteratively, such that it is compatible to the stochastic gradient descent algorithm as follows:}
\begin{equation}
 \mathcal{L}_{4} = \max_{f}\left(\text{ReLU}\left(\frac{n_f}{m_f}-1\right)\right).
 \label{eq:nmr}
\end{equation}
\zhenk{
The rectified linear units (ReLU) function excludes the contribution of the inaudible noise to the loss when $n_f/m_f -1 < 0$. Out of those frequency bins where the noise is audible, the $\max$ operator selects the one with the largest NMR, which counts towards the total loss. The process as such resembles MP3's bit allocation algorithm, as it tackles the frequency bin with the largest NMR for each training iteration.}

\section{Experiments}
\label{sec:exp}

\subsection{Experimental Setup} 

\subsubsection{Data Preparation and Hyperparameters}
Our training dataset consists of 1,000 single-channel clips of commercial music, spanning 13 genres. Each clip is about 20 seconds long, amounting to about 5.5 hours of play time. The sampling rate is 44.1 kHz and downsampled to 32 kHz for the lower bitrate setup. Each frame contains $T=512$ samples with an overlap of $32$ samples, where a Hann window is applied to the overlapping region. Note that the choice of frame size is to align the system's hyperparameters to the previous work \cite{KankanahalliS2018icassp, zhen2020efficient, ZhenK2019interspeech}, but it does not necessarily mean that 512 results in an enough frequency resolution for PAM-based lost terms. For training, hyperparameters are found based on validation with another 104 clips: 128 frames for the batch size; $\alpha=300$ for the initial softmax scaling factor; $2\times 10^{-4}$ for the initial learning rate of the Adam optimizer \cite{KingmaD2015adam}, and $2\times10^{-5}$ for the second cascaded modules; 64 and 32 kernels for the quantization for low and high bitrate cases, respectively; 50 and 30 for the number of epochs to train the first and the second modules in CMRL, respectively.

\begin{figure}[t]
   \centering
   \subfigure[Low bitrates]{\includegraphics[height=2.7in]{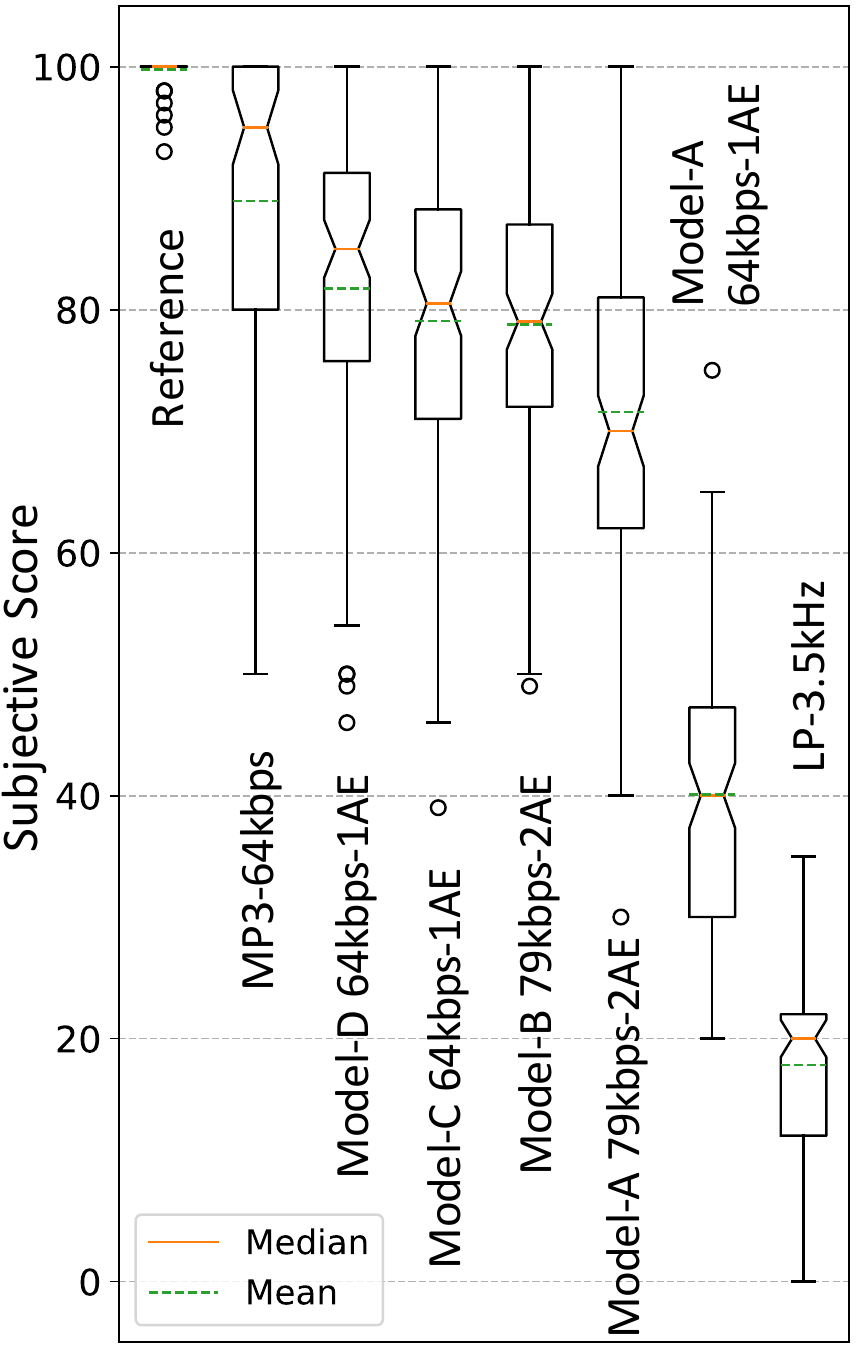}} 
  ~~
  \subfigure[High bitrates]{\includegraphics[height=2.7in]{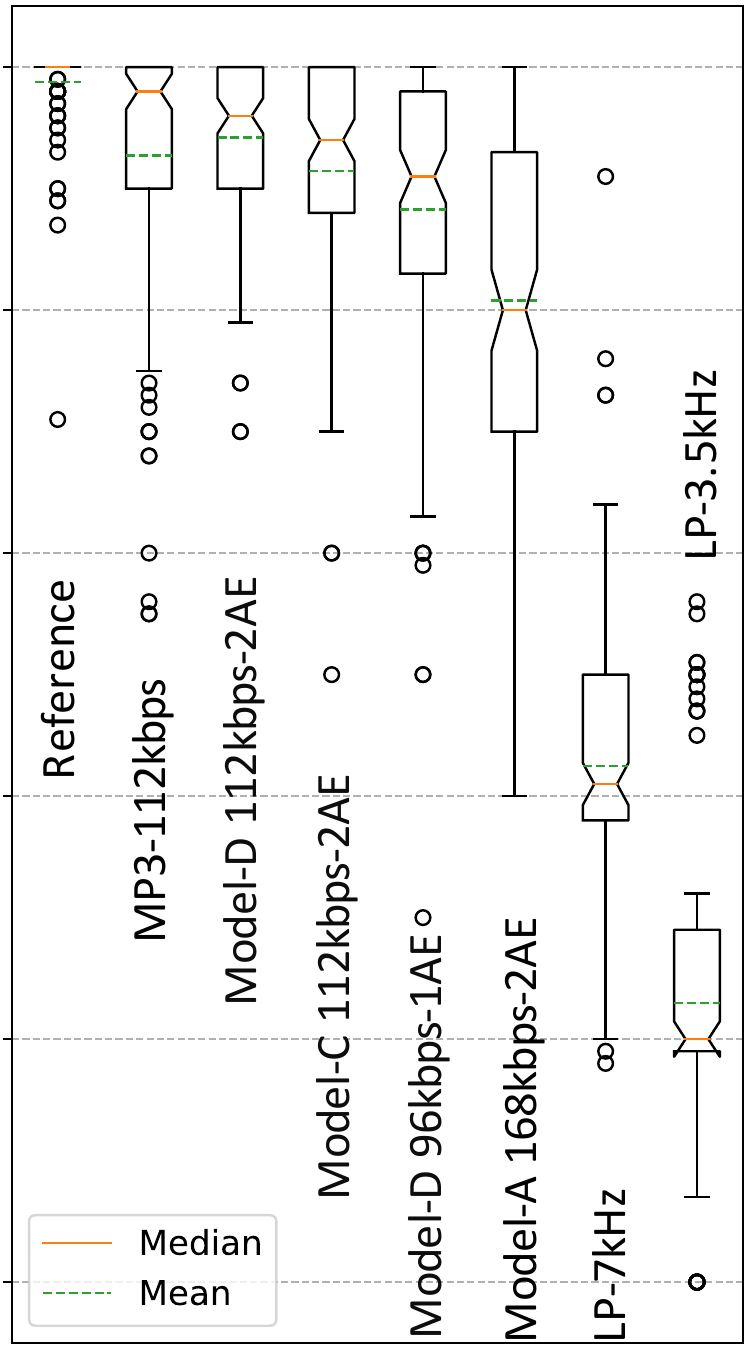}} 
  \vspace{-0.09in}
    \caption{Subjective scores from the MUSHRA tests.}
    \vspace{-0.09in}
    \label{fig:mushra}
\end{figure}

\subsubsection{Competing Models}
%

We consider two baseline models: Model-A trained by the SSE ($\mathcal{L}_1$) and Model-B also learns on the mel-frequency loss ($\mathcal{L}_2$) as in recent neural speech codecs \cite{KankanahalliS2018icassp, ZhenK2019interspeech, zhen2020efficient}. We validate the priority weighting loss ($\mathcal{L}_3$) in Model-C, and noise modulation ($\mathcal{L}_4$) in Model-D.
All models are based on the architecture discussed in Section.\ref{sec:nac}.
\begin{align}
\nonumber\calL&=\calL_1~\text{: Model-A,}\quad \calL=\calL_1+\lambda\calL_2~&&\text{: Model-B,}\\
\nonumber\calL&=\calL_1+\lambda(\calL_2+\calL_3)~&&\text{: Model-C,}\\
\nonumber\calL&=\calL_1+\lambda(\calL_2+\calL_3+\calL_4)~&&\text{: Model-D.}
\end{align}
\zhenk{Note that the scale for the time-domain SSE $\mathcal{L}_1$ differs from the other frequency-domain loss terms. 
Adapting from the setup in \cite{KankanahalliS2018icassp}, we find that simply choosing one blending weight $\lambda=0.1$ for $\calL_2$, $\calL_3$ and $\calL_4$ shows good results. }

Each model is also specified by the target bitrate and model complexity, e.g.,  ``Model-A 168kbps-2AE" is equipped with two concatenated AEs, trained by $\mathcal{L}_1$ for a bitrate of 168 kbps.

\subsection{Experimental Results}
Ten audio experts participated our two MUSHRA listening tests for low and high bitrate settings using headphones. We post-screened one of them as per the guideline \cite{mushra}. We randomly sample 13 songs, one per genre, and fix them throughout all tests. Fig. \ref{fig:mushra} summarizes the test results. \zhenk{Each box extends from the lower to the upper quartile with a $95\%$ confidence interval (the notch) of the median (the orange hard line). The mean scores and outliers are also shown in the green dotted line and circles, respectively.}

\begin{figure}[t]
   \centering
   \subfigure[No noise modulation (Model-C). Noise can exceed the mask (orange).]{\includegraphics[width=.99\columnwidth]{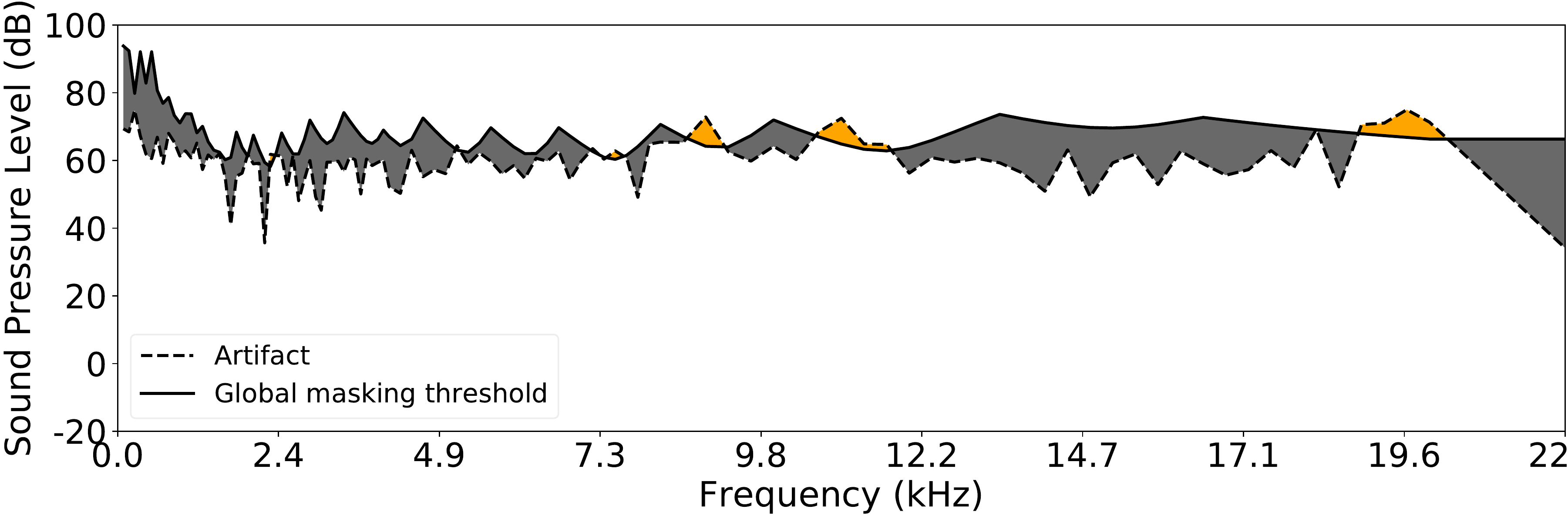}} 
   \subfigure[With noise modulation (Model-D)]{\hfill\includegraphics[width=.99\columnwidth]{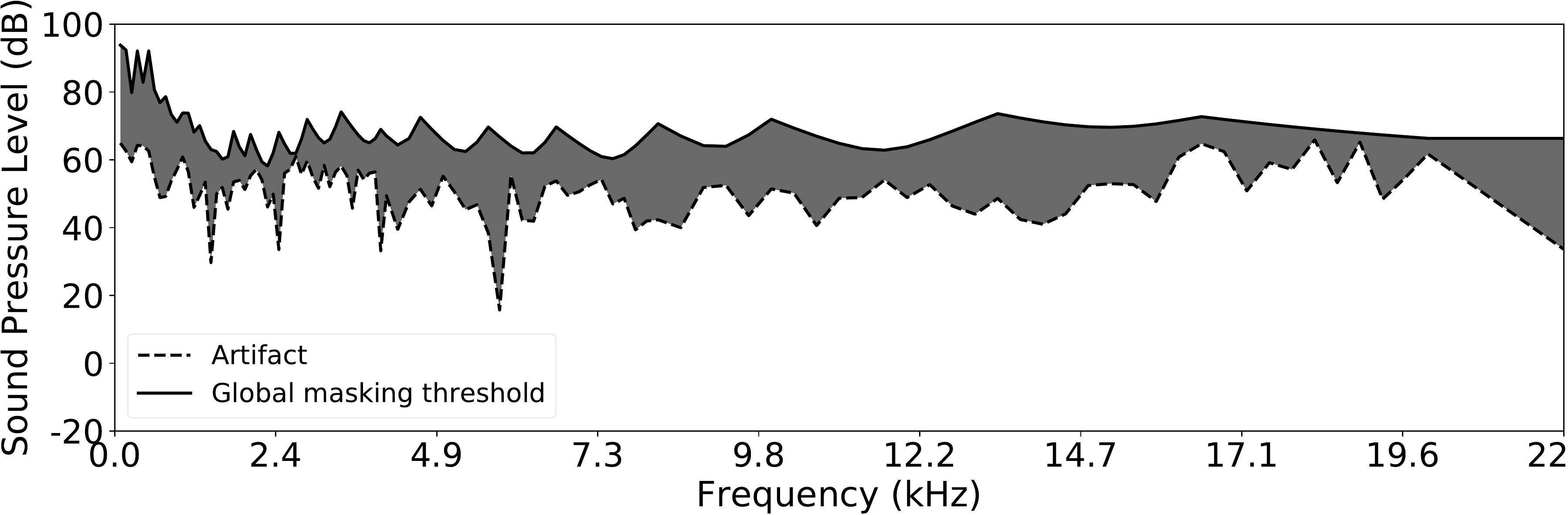}} 
   \vspace{-0.09in}
    \caption{The effect of the proposed noise modulation loss.}
    \vspace{-0.09in}
    \label{fig:MNR}
\end{figure}

The low bitrate session targets at 64 kbps with the sample rate of 32 kHz. 
As illustrated in Fig. \ref{fig:mushra} (a), the baseline trained purely on SSE does not perform well (Model-A). However, the additional loss term $\mathcal{L}_2$ defined in the mel-scaled frequency domain improves the performance (Model-B 79kbps 2AE$>$ Model-A 79kbps 2AE). Finally, 
via psychoacoustic weighting, Model-C with a smaller architecture and lower bitrate (i.e., one AE at 64 kbps) is on par with Model-B with two AEs at 79 kbps.
Model-D with both PAM-inspired loss terms receives the highest subjective score among the NAC systems, which justifies the effectiveness of the proposed method with a noticeable coding gain. 
However, since our codec is lightweight as a neural network, Model-D does not outperform the commercial MP3 codec from Adobe Audition\textregistered ~(licensed from Fraunhofer IIS and Thomson). We expect that a more complex model can catch up with this gap by employing the proposed psychoacoustic loss functions.

The high bitrate session includes the hidden reference, two anchors (filtered at 3.5 kHz and 7 kHz), the commercial MP3 codec at 112 kbps, and four NAC systems, sampled at 44.1 kHz.
In Fig. \ref{fig:mushra} (b), with both priority weighting and noise modulation, Model-D outperforms Model-A which is twice as large and performs at a $64.3\%$ higher bitrate. Model-D is also superior to Model-C at the same bitrate and model complexity thanks to noise modulation. With 900K parameters, Model-D achieves almost transparent quality similar to MP3 at the same bitrate (Model-D 112kbps-2AE vs. MP3-112 kbps).

The superiority of Model-D over Model-C is additionally explained in Fig. \ref{fig:MNR}.
While Model-C can result in audible reconstruction error (organge regions in (a)), the noise modulation loss ($\calL_4$) in Model-D suppresses it under the masking curve in (b), leading to a higher perceptual quality.

\section{Conclusion}
\label{sec:con}
We showed that incorporating the simultaneous masking effect in the objective function is advantageous to NAC in terms of the coding gain and model efficiency. Although the system is based on PAM-1, it successfully proved the concept and suggests that a more advanced PAM, e.g., by employing temporal masking, will improve the performance further. We also publicized all source codes and demo signals\footnote{Available at 
\href{https://saige.sice.indiana.edu/research-projects/pam-nac}{https://saige.sice.indiana.edu/research-projects/pam-nac}}.

\bibliographystyle{IEEEtran}
\bibliography{mjkim}

\begin{thebibliography}{10}
\providecommand{\url}[1]{#1}
\csname url@samestyle\endcsname
\providecommand{\newblock}{\relax}
\providecommand{\bibinfo}[2]{#2}
\providecommand{\BIBentrySTDinterwordspacing}{\spaceskip=0pt\relax}
\providecommand{\BIBentryALTinterwordstretchfactor}{4}
\providecommand{\BIBentryALTinterwordspacing}{\spaceskip=\fontdimen2\font plus
\BIBentryALTinterwordstretchfactor\fontdimen3\font minus
  \fontdimen4\font\relax}
\providecommand{\BIBforeignlanguage}[2]{{%
\expandafter\ifx\csname l@#1\endcsname\relax
\typeout{** WARNING: IEEEtran.bst: No hyphenation pattern has been}%
\typeout{** loaded for the language `#1'. Using the pattern for}%
\typeout{** the default language instead.}%
\else
\language=\csname l@#1\endcsname
\fi
#2}}
\providecommand{\BIBdecl}{\relax}
\BIBdecl

\bibitem{BosiM1997iso}
M.~Bosi, K.~Brandenburg, S.~Quackenbush, L.~Fielder, K.~Akagiri, H.~Fuchs, and
  M.~Dietz, ``{ISO/IEC MPEG-2} advanced audio coding,'' \emph{Journal of the
  Audio Engineering Society}, vol.~45, no.~10, pp. 789--814, 1997.

\bibitem{mp3}
{{ISO/IEC} 11172-3:1993}, ``Coding of moving pictures and associated audio for
  digital storage media at up to about 1.5 mbit/s,'' 1993.

\bibitem{BrandenburgK1994iso}
K.~Brandenburg and G.~Stoll, ``{ISO/MPEG-1 audio:} a generic standard for
  coding of high-quality digital audio,'' \emph{Journal of the Audio
  Engineering Society}, vol.~42, no.~10, pp. 780--792, 1994.

\bibitem{engel2017neural}
J.~Engel, C.~Resnick, A.~Roberts, S.~Dieleman, M.~Norouzi, D.~Eck, and
  K.~Simonyan, ``Neural audio synthesis of musical notes with wave{N}et
  autoencoders,'' in \emph{Proceedings of the 34th International Conference on
  Machine Learning-Volume 70}.\hskip 1em plus 0.5em minus 0.4em\relax JMLR.
  org, 2017, pp. 1068--1077.

\bibitem{OordA2016wavenet}
A.~{van den Oord}, S.~Dieleman, H.~Zen, K.~Simonyan, O.~Vinyals, A.~Graves,
  N.~Kalchbrenner, A.~Senior, and K.~Kavukcuoglu, ``Wave{N}et: A generative
  model for raw audio,'' \emph{arXiv preprint arXiv:1609.03499}, 2016.

\bibitem{KlejsaJ2019samplernn}
J.~Klejsa, P.~Hedelin, C.~Zhou, R.~Fejgin, and L.~Villemoes, ``High-quality
  speech coding with {SampleRNN},'' in \emph{Proceedings of the IEEE
  International Conference on Acoustics, Speech, and Signal Processing
  (ICASSP)}, 2019.

\bibitem{StollerD2018waveunet}
D.~Stoller, S.~Ewert, and S.~Dixon, ``Wave-{U}-{N}et: A multi-scale neural
  network for end-to-end audio source separation,'' in \emph{Proceedings of the
  International Conference on Music Information Retrieval (ISMIR)}, 2018.

\bibitem{GarbaceaC2019vqvae}
Y.~L. C.~Garbacea, A.~{van den Oord}, ``Low bit-rate speech coding with
  {VQ-VAE} and a wave{N}et decoder,'' in \emph{Proceedings of the IEEE
  International Conference on Acoustics, Speech, and Signal Processing
  (ICASSP)}, 2019.

\bibitem{OordA2017vqvae}
A.~{van den Oord}, O.~Vinyals, and K.~Kavukcuoglu, ``Neural discrete
  representation learning,'' in \emph{Advances in Neural Information Processing
  Systems (NIPS)}, 2017, pp. 6306--6315.

\bibitem{ValinJ2019lpcnet}
J.-M. Valin and J.~Skoglund, ``{LPCNet}: Improving neural speech synthesis
  through linear prediction,'' in \emph{Proceedings of the IEEE International
  Conference on Acoustics, Speech, and Signal Processing (ICASSP)}, 2019.

\bibitem{ChenZ2017deep}
Z.~Chen, Y.~Luo, and N.~Mesgarani, ``Deep attractor network for
  single-microphone speaker separation,'' in \emph{Acoustics, Speech and Signal
  Processing (ICASSP), 2017 IEEE International Conference on}.\hskip 1em plus
  0.5em minus 0.4em\relax IEEE, 2017, pp. 246--250.

\bibitem{LiuQ2017perceptually}
Q.~J. Liu, W.~W. Wang, P.~J. Jackson, and Y.~Tang, ``A perceptually-weighted
  deep neural network for monaural speech enhancement in various background
  noise conditions,'' in \emph{2017 25th European Signal Processing Conference
  (EUSIPCO)}.\hskip 1em plus 0.5em minus 0.4em\relax IEEE, 2017, pp.
  1270--1274.

\bibitem{zhao2019perceptual}
Z.~Y. Zhao, S.~Elshamy, and T.~Fingscheidt, ``A perceptual weighting filter
  loss for {DNN} training in speech enhancement,'' in \emph{2019 IEEE Workshop
  on Applications of Signal Processing to Audio and Acoustics (WASPAA)}.\hskip
  1em plus 0.5em minus 0.4em\relax IEEE, 2019, pp. 229--233.

\bibitem{kumar2016speech}
A.~Kumar and D.~Florencio, ``Speech enhancement in multiple-noise conditions
  using deep neural networks,'' in \emph{Proceedings of the Annual Conference
  of the International Speech Communication Association (Interspeech)}, San
  Francisco, CA, USA, 2016, pp. 352--356.

\bibitem{ZhenK2018psychoacoustically}
K.~Zhen, A.~Sivaraman, J.~M. Sung, and M.~Kim, ``On psychoacoustically weighted
  cost functions towards resource-efficient deep neural networks for speech
  denoising,'' \emph{arXiv preprint arXiv:1801.09774}, 2018.

\bibitem{TaalC2010icassp}
C.~H. Taal, R.~C. Hendriks, R.~Heusdens, and J.~Jensen, ``A short-time
  objective intelligibility measure for time-frequency weighted noisy speech,''
  in \emph{Proceedings of the IEEE International Conference on Acoustics,
  Speech, and Signal Processing (ICASSP)}, 2010.

\bibitem{RixA2001pesq}
A.~W. Rix, J.~G. Beerends, M.~P. Hollier, and A.~P. Hekstra, ``Perceptual
  evaluation of speech quality ({PESQ})-a new method for speech quality
  assessment of telephone networks and codecs,'' in \emph{Proceedings of the
  IEEE International Conference on Acoustics, Speech, and Signal Processing
  (ICASSP)}, vol.~2, 2001, pp. 749--752.

\bibitem{FuS2018e2e}
S.~W. Fu, T.~W. Wang, Y.~Tsao, X.~G. Lu, and H.~Kawai, ``End-to-end waveform
  utterance enhancement for direct evaluation metrics optimization by fully
  convolutional neural networks,'' \emph{IEEE/ACM Transactions on Audio,
  Speech, and Language Processing}, vol.~26, no.~9, pp. 1570--1584, 2018.

\bibitem{MartinJ2018deep}
J.~M. Mart{\'\i}n-Do{\~n}as, A.~M. Gomez, J.~A. Gonzalez, and A.~M. Peinado,
  ``A deep learning loss function based on the perceptual evaluation of the
  speech quality,'' \emph{IEEE Signal processing letters}, vol.~25, no.~11, pp.
  1680--1684, 2018.

\bibitem{PainterT2000ieeeproc}
T.~Painter and A.~Spanias, ``Perceptual coding of digital audio,''
  \emph{Proceedings of the IEEE}, vol.~88, no.~4, pp. 451--515, 2000.

\bibitem{KankanahalliS2018icassp}
S.~Kankanahalli, ``End-to-end optimized speech coding with deep neural
  networks,'' in \emph{Proceedings of the IEEE International Conference on
  Acoustics, Speech, and Signal Processing (ICASSP)}, 2018.

\bibitem{TanK2019taslp}
K.~Tan, J.~Chen, and D.~Wang, ``Gated residual networks with dilated
  convolutions for monaural speech enhancement,'' \emph{IEEE/ACM Transactions
  on Audio, Speech, and Language Processing}, vol.~27, pp. 189--198, 2019.

\bibitem{HeK2016cvpr}
K.~He, X.~Zhang, S.~Ren, and J.~Sun, ``Deep residual learning for image
  recognition,'' in \emph{Proceedings of the IEEE International Conference on
  Computer Vision and Pattern Recognition (CVPR)}, 2016, pp. 770--778.

\bibitem{AgustssonE2017softmax}
E.~Agustsson, F.~Mentzer, M.~Tschannen, L.~Cavigelli, R.~Timofte, L.~Benini,
  and L.~V. Gool, ``Soft-to-hard vector quantization for end-to-end learning
  compressible representations,'' in \emph{Advances in Neural Information
  Processing Systems (NIPS)}, 2017, pp. 1141--1151.

\bibitem{ShiW2016superresolution}
W.~Shi, J.~Caballero, F.~Husz{\'a}r, J.~Totz, A.~P. Aitken, R.~Bishop,
  D.~Rueckert, and Z.~Wang, ``Real-time single image and video super-resolution
  using an efficient sub-pixel convolutional neural network,'' in
  \emph{Proceedings of the IEEE International Conference on Computer Vision and
  Pattern Recognition (CVPR)}, 2016, pp. 1874--1883.

\bibitem{ZhenK2019interspeech}
K.~Zhen, J.~Sung, M.~S. Lee, S.~Beack, and M.~Kim, ``Cascaded cross-module
  residual learning towards lightweight end-to-end speech coding,'' in
  \emph{Proceedings of the Annual Conference of the International Speech
  Communication Association (Interspeech)}, 2019.

\bibitem{GershoA1983vq}
A.~Gersho and V.~Cuperman, ``Vector quantization: A pattern-matching technique
  for speech coding,'' \emph{IEEE Communications magazine}, vol.~21, no.~9, pp.
  15--21, 1983.

\bibitem{salomon2004data}
D.~Salomon, \emph{Data compression: the complete reference}.\hskip 1em plus
  0.5em minus 0.4em\relax Springer Science \& Business Media, 2004.

\bibitem{yen2005low}
C.~H. Yen, Y.~S. Lin, and B.~F. Wu, ``A low-complexity {MP}3 algorithm that
  uses a new rate control and a fast dequantization,'' \emph{IEEE Transactions
  on Consumer Electronics}, vol.~51, no.~2, pp. 571--579, 2005.

\bibitem{zamani2019spatial}
S.~Zamani and K.~Rose, ``Spatial audio coding without recourse to background
  signal compression,'' in \emph{ICASSP 2019-2019 IEEE International Conference
  on Acoustics, Speech and Signal Processing (ICASSP)}.\hskip 1em plus 0.5em
  minus 0.4em\relax IEEE, 2019, pp. 720--724.

\bibitem{zhen2020efficient}
K.~Zhen, M.~S. Lee, J.~Sung, S.~Beack, and M.~Kim, ``Efficient and scalable
  neural residual waveform coding with collaborative quantization,'' in
  \emph{ICASSP 2020-2020 IEEE International Conference on Acoustics, Speech and
  Signal Processing (ICASSP)}.\hskip 1em plus 0.5em minus 0.4em\relax IEEE,
  2020, pp. 361--365.

\bibitem{KingmaD2015adam}
D.~P. Kingma and J.~Ba, ``Adam: A method for stochastic optimization,'' in
  \emph{Proceedings of the International Conference on Learning Representations
  (ICLR)}, 2015.

\bibitem{mushra}
{{ITU-R} {Recommendation} {BS} 1534-1}, ``Method for the subjective assessment
  of intermediate quality levels of coding systems ({MUSHRA}),'' 2003.

\end{thebibliography}

\end{document}